\documentclass[pdftex,twocolumn]{aastex6}

\usepackage{apjfonts,graphicx,graphics}

\usepackage{graphicx}
\usepackage{natbib}
\usepackage{mathptmx}
\usepackage{amsmath}
\usepackage{wasysym}
\usepackage{times}
\usepackage{epsfig}
\usepackage{ulem}

\newcommand{\Mpc}{\rm\; Mpc}
\newcommand{\kpc}{\rm\; kpc}

\newcommand{\km}{\rm\; km}
\newcommand{\m}{\rm\; m}
\newcommand{\cm}{\rm\; cm}

\newcommand{\yr}{\rm\; yr}

\newcommand{\Myr}{\rm\; Myr}
\newcommand{\s}{\rm\; s}

\newcommand{\GHz}{\rm\; GHz}

\newcommand{\kHz}{\rm\; kHz}

\newcommand{\K}{\rm\; K}

\newcommand{\Msun}{\hbox{$\rm\thinspace M_{\odot}$}}

\newcommand{\Msunpcukpc}{\hbox{$\Msun\kpc^{-3}\,$}}

\newcommand{\Msunpyr}{\hbox{$\Msun\yr^{-1}\,$}}
\newcommand{\Msunpyrpsqkpc}{\hbox{$\Msunpyr\kpc^{-2}\,$}}

\newcommand{\erg}{\rm\; erg}
\newcommand{\Jy}{\rm\; Jy}
\newcommand{\mJy}{\rm\; mJy}

\newcommand{\ergps}{\hbox{$\erg\s^{-1}\,$}}

\newcommand{\Jykmps}{\hbox{$\Jy\km\s^{-1}\,$}}
\newcommand{\Jypbmkmps}{\hbox{$\Jy{\rm /beam}.{\rm km}\s^{-1}\,$}}
\newcommand{\mJypbm}{\hbox{$\mJy{\rm /beam}$}}

\newcommand{\kmps}{\hbox{$\km\s^{-1}\,$}}

\newcommand{\kmpspMpc}{\hbox{$\kmps\Mpc^{-1}\,$}}

\newcommand{\Zsun}{\hbox{$\thinspace \mathrm{Z}_{\odot}$}}

\newcommand{\asec}{\rm\; arcsec}

\newcommand{\psqcm}{\hbox{$\cm^{-2}\,$}}

\newcommand{\COtoH}{\hbox{$\psqcm(\K\kmps)^{-1}$}}

\begin{document}


\title{ALMA observations of massive molecular gas filaments encasing radio bubbles in the Phoenix cluster}
\author{H.~R. Russell$^{1*}$, M. McDonald$^{2}$, B.~R. McNamara$^{3,4}$, A.~C. Fabian$^{1}$, P.~E.~J. Nulsen$^{5,6}$, M.~B. Bayliss$^{2,7}$, B.~A. Benson$^{8,9,10}$, M. Brodwin$^{11}$, J.~E. Carlstrom$^{10,9}$, A.~C. Edge$^{12}$, J. Hlavacek-Larrondo$^{13}$, D.~P. Marrone$^{14}$, C.~L. Reichardt$^{15}$ and J.~D. Vieira$^{16}$}
\affil{$^1$ Institute of Astronomy, Madingley Road, Cambridge CB3 0HA\\
    $^2$ Kavli Institute for Astrophysics and Space Research, Massachusetts Institute of Technology, 77 Massachusetts Avenue, Cambridge, MA 02139, USA\\
    $^3$ Department of Physics and Astronomy, University of Waterloo, Waterloo, ON N2L 3G1, Canada\\
    $^4$ Perimeter Institute for Theoretical Physics, Waterloo, ON N2L 2Y5, Canada\\
    $^{5}$Harvard-Smithsonian Center for Astrophysics, 60 Garden Street, Cambridge, MA 02138, USA\\
    $^{6}$ICRAR, University of Western Australia, 35 Stirling Hwy, Crawley, WA 6009, Australia\\
    $^{7}$Department of Physics and Astronomy, Colby College, 5100 Mayflower Hill Dr, Waterville, ME 04901, USA\\
    $^{8}$Fermi National Accelerator Laboratory, Batavia, IL 60510-0500, USA\\
    $^{9}$Department of Astronomy and Astrophysics, University of Chicago, Chicago, IL 60637, USA\\
    $^{10}$Kavli Institute for Cosmological Physics, University of Chicago, Chicago, IL 60637, USA\\
    $^{11}$Department of Physics and Astronomy, University of Missouri, Kansas City, MO 64110, USA\\
    $^{12}$Department of Physics, Durham University, Durham DH1 3LE\\
    $^{13}$D\'epartement de Physique, Universit\'e de Montr\'eal, Montr\'eal, QC H3C 3J7, Canada\\
    $^{14}$Steward Observatory, University of Arizona, 933 North Cherry Avenue, Tucson, AZ 85721, USA\\
    $^{15}$School of Physics, University of Melbourne, Parkville VIC 3010, Australia\\
    $^{16}$Department of Astronomy and Department of Physics, University of Illinois, 1002 West Green St., Urbana, IL 61801, USA} 

\altaffiltext{$^*$}{hrr27@ast.cam.ac.uk}

\begin{abstract}
We report new ALMA observations of the CO(3-2) line emission from the $2.1\pm0.3\times10^{10}\Msun$ molecular gas reservoir in the central galaxy of the Phoenix cluster.  The cold molecular gas is fuelling a vigorous starburst at a rate of $500-800\Msunpyr$ and powerful black hole activity in the form of both intense quasar radiation and radio jets.  The radio jets have inflated huge bubbles filled with relativistic plasma into the hot, X-ray atmospheres surrounding the host galaxy.  The ALMA observations show that extended filaments of molecular gas, each $10-20\kpc$ long with a mass of several billion solar masses, are located along the peripheries of the radio bubbles.  The smooth velocity gradients and narrow line widths along each filament reveal massive, ordered molecular gas flows around each bubble, which are inconsistent with gravitational free-fall.  The molecular clouds have been lifted directly by the radio bubbles, or formed via thermal instabilities induced in low entropy gas lifted in the updraft of the bubbles. These new data provide compelling evidence for close coupling between the radio bubbles and the cold gas, which is essential to explain the self-regulation of feedback.  The very feedback mechanism that heats hot atmospheres and suppresses star formation may also paradoxically stimulate production of the cold gas required to sustain feedback in massive galaxies.
\end{abstract}



\section{Introduction}
\label{sec:intro}

The energy output by active galactic nuclei (AGN) has long been
recognized as sufficient to unbind the interstellar medium from even
the most massive host galaxies (\citealt{Silk98}).  Recent
observations of ionized and molecular gas outflows driven by intense
radiation or radio jet activity from the central AGN show that this
energy can be efficiently coupled to the surrounding interstellar gas
(eg. \citealt{Morganti05}; \citealt{Nesvadba06}; \citealt{Feruglio10};
\citealt{Rupke11}; \citealt{Alatalo11}; \citealt{Dasyra11};
\citealt{Morganti15}).  \textit{Chandra} X-ray observations of the hot
atmospheres surrounding giant elliptical galaxies and central cluster
galaxies have also revealed huge cavities where the hot gas has been
displaced by expanding radio bubbles inflated by radio jets
(\citealt{McNamara00}; \citealt{FabianPer00,FabianPer06}).  Known as
AGN feedback, these energetic outbursts are therefore observed to
couple effectively to the cold and warm interstellar gas and the hot
gas atmospheres surrounding massive galaxies.  AGN feedback is an
essential mechanism in galaxy formation that powers gas outflows to
truncate massive galaxy growth.  This process is thought to produce
the observed evolution of galaxies from gas-rich, star forming systems
to `red and dead' ellipticals and imprint the observed coevolution of
massive galaxies and supermassive black holes (SMBHs;
\citealt{Magorrian98,Croton06,Bower06}).


However, the details of how a SMBH can regulate the growth of its host
environment over many orders of magnitude in spatial scale are still
poorly understood.  In the most massive galaxies at the centres of
cool core galaxy clusters, the radiative cooling time of the hot gas
atmosphere can fall below a Gyr and heat input from the AGN must be
distributed throughout the central $100\kpc$ to prevent the formation
of a cooling flow (eg. \citealt{Edge92}; \citealt{Peres98};
\citealt{VoigtFabian04}).  Without this energy input, gas would cool
unimpeded from the cluster atmosphere and produce at least an order of
magnitude more molecular gas and star formation than is observed in
central cluster galaxies (\citealt{Johnstone87,Edge01,Salome03}).
Radio jets powered by the central AGN inflate buoyant radio bubbles
and drive shocks and sound waves into the intracluster medium to
produce distributed heating throughout the cluster core (for reviews
see \citealt{McNamaraNulsen07,McNamara12,Fabian12}).  X-ray studies of
large samples of galaxy groups and clusters show that this energy
input is sufficient to replace the majority of the radiative losses
from the cluster gas on large scales
(\citealt{Birzan04,DunnFabian06,Rafferty06}).  The heating rate
supplied by the AGN is also observed to be closely correlated with the
cooling rate of the cluster atmosphere, which implies a highly
effective feedback loop operating over this huge range of spatial
scales.  A few per cent of the most rapidly cooling cluster gas does
cool to low temperatures and likely feeds the observed cold molecular
gas reservoirs and star formation in the central galaxy.  Although the
level of gas cooling falls far below the predictions of cooling flows,
prompt accretion of this residual component is likely required to link
the large scale cooling rate to the energy output of the AGN in an
efficient feedback loop.

Observations of ionized and molecular gas at the centres of clusters
have revealed cool gas filaments extending radially from the galaxy
centre towards radio bubbles inflated by the jet
(\citealt{FabianFil03}; \citealt{Salome06,Salome08};
\citealt{HatchPer06}; \citealt{Lim08}).  In the Perseus cluster, the
velocity structure of the H$\alpha$-emitting filaments, which are
coincident with detections of CO emission from the IRAM $30\m$
telescope, traces streamlines underneath a buoyantly rising radio
bubble (\citealt{Salome06,Salome11}).  ALMA observations of molecular
gas at the centres of clusters
(\citealt{Russell14,Russell16,McNamara14,David14,Tremblay16,Vantyghem16}) have
shown cold gas filaments extending along the trajectories of radio
bubbles.  The molecular clouds have either been lifted directly by the
bubbles or cooled in situ from warmer, thermally unstable gas lifted
in their wakes.  The velocities of the molecular clouds are remarkably
slow compared to the stellar velocity dispersion in these massive
galaxies and lie well below the galaxy's escape velocity.  The
molecular gas will likely fall back towards the galaxy centre and fuel
both star formation and future AGN activity.

Here we present new ALMA observations of the CO(3-2) emission from the
molecular gas in the central galaxy of the Phoenix cluster.
Discovered with the South Pole Telescope, the Phoenix cluster
(SPT-CLJ2344-4243), at redshift $z=0.596$, is the most luminous X-ray
cluster known (\citealt{Williamson11,McDonald12}), and the
$500-800\Msunpyr$ starburst hosted by its central galaxy is amongst
the largest found in any galaxy below redshift 1.  The star formation
is observed in bright filaments stretching beyond $100\kpc$, sustained
by a 20 billion solar mass reservoir of molecular gas
(\citealt{McDonald13,McDonald14}).  The stellar mass of the massive
central galaxy is $3\times10^{12}\Msun$
(\citealt{McDonald12,McDonald13}) and it hosts an unusual central
supermassive black hole that is powering both intense radiation and
relativistic jets.  Observations show these to be distinct modes of
AGN feedback.  The black hole may be in the process of transitioning
from a radiatively powerful quasar to a radio galaxy (eg. \citealt{Churazov05,Russell13,HlavacekLarrondo13}) whose mechanical
power output of $\sim10^{46}\ergps$ is among the largest measured
(eg. \citealt{HlavacekLarrondo15,McDonald15}).  The Phoenix cluster
therefore hosts an extreme example of this common mechanism in galaxy
evolution.  Both the powerful black hole activity and the vigorous
starburst are fuelled by the massive cold molecular gas reservoir,
whose structure can now be resolved with ALMA to understand how these
processes are regulated.

We assume a standard $\Lambda$CDM cosmology with $H_{0}=70\kmpspMpc$,
$\Omega_{\mathrm{M}}=0.27$ and $\Omega_{\Lambda}=0.73$.  At the
redshift of the Phoenix cluster $z=0.596$ (\citealt{Ruel14};
\citealt{McDonald15}), 1 arcsec corresponds to $6.75\kpc$.  


\section{Data reduction}

\begin{figure*}
\begin{minipage}{\textwidth}
\centering
\includegraphics[width=0.53\columnwidth]{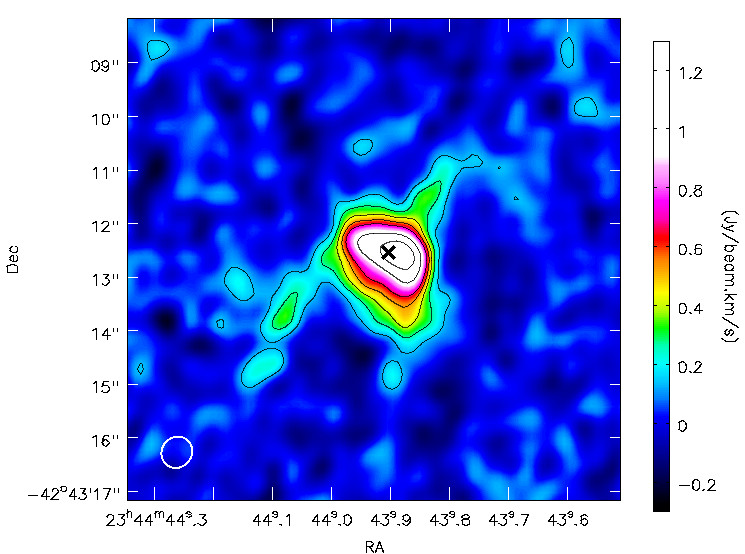}
\raisebox{0.5cm}{\includegraphics[width=0.45\columnwidth]{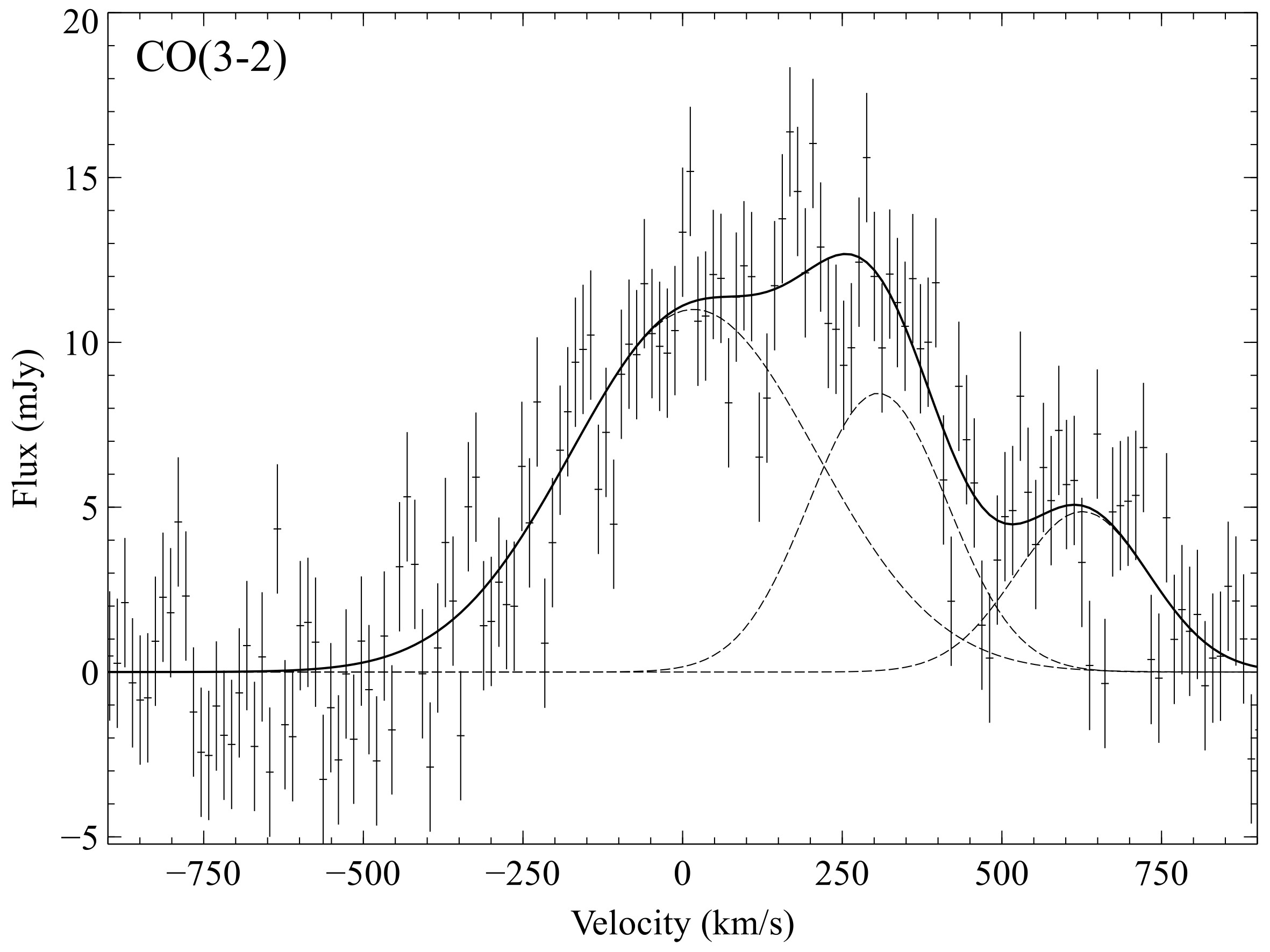}}
\caption{Left: Phoenix CO(3-2) integrated intensity map for velocities $-430$ to $+600\kmps$ covering both the central gas peak and the extended filaments.  Contour levels are $2\sigma$, $4\sigma$, $6\sigma$, $8\sigma$, $10\sigma$, $15\sigma$ etc, where $\sigma=0.067\Jypbmkmps$.  The ALMA beam is shown lower left and the continuum point source location is shown by the black cross.  Right: Phoenix CO(3-2) spectrum for a $6\arcsec\times6\arcsec$ region centred on the nucleus.  The best fit model is shown by the solid black line and individual Gaussian components are shown by the dashed lines (see Table \ref{tab:fits}).}
\label{fig:totalspec}
\end{minipage}
\end{figure*}

\begin{figure*}
\begin{minipage}{\textwidth}
\includegraphics[width=0.56\columnwidth]{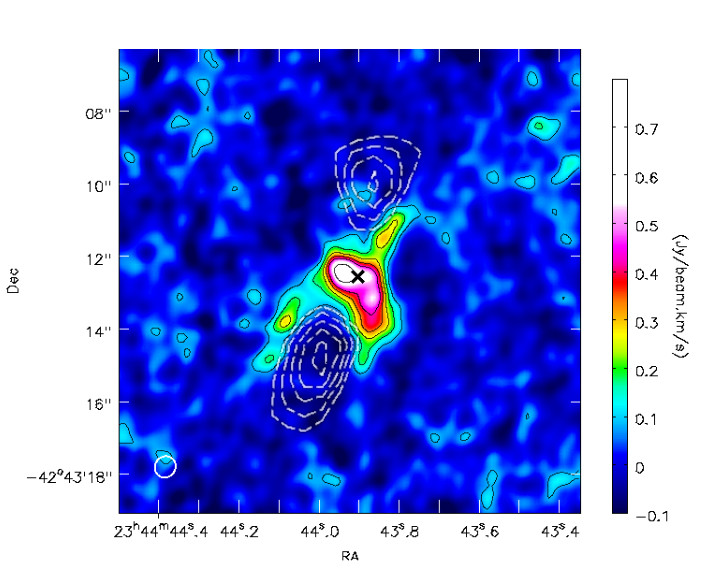}
\raisebox{0.65cm}{\includegraphics[width=0.38\columnwidth]{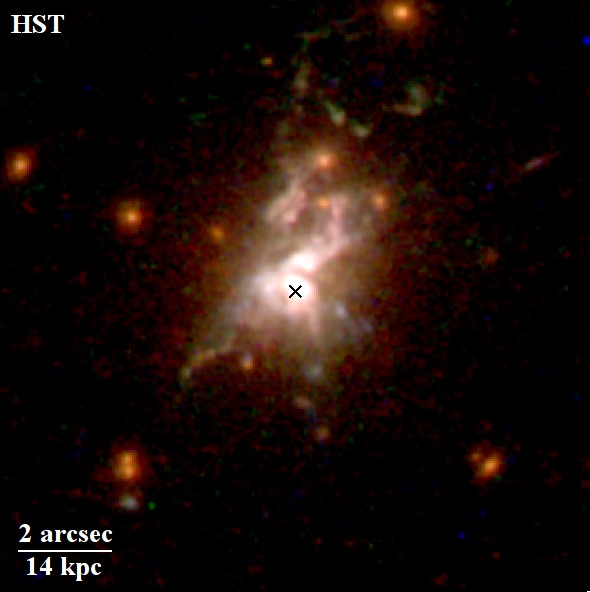}}
\caption{Left: CO(3-2) integrated intensity map for velocities $0$ to $+480\kmps$ covering the extended filaments.  Contour levels are at $2\sigma$, $4\sigma$, $6\sigma$, $8\sigma$, $10\sigma$, $15\sigma$, $20\sigma$ etc., where $\sigma=0.042\Jypbmkmps$.  The ALMA beam is shown in the lower left corner ($0.60\asec\times0.56\asec$).  The X-ray cavities are shown by the dashed white contours, which correspond to the negative residuals after a smooth model was subtracted from the X-ray surface brightness (\citealt{McDonald15}).  Right: HST image combining the F475W (blue), F625W (green) and F814W (red) filters (\citealt{McDonald13}).  Both images cover the same field of view.}
\label{fig:filimg}
\end{minipage}
\end{figure*}

The brightest cluster galaxy (BCG) in the Phoenix cluster was observed
by ALMA on 15 and 16 June 2014 (Cycle 2, ID 2013.1.01302.S; PI
McDonald) simultaneously covering the CO(3-2) line at $216.66\GHz$ and
the sub-mm continuum emission in three additional basebands at 219.5,
230.5 and $232.5\GHz$.  The single pointing observations were centred
on the nucleus with a field of view of $28.5\asec$.  The total time on
source was $58.5\min$ split into nine $\sim6\min$ observations and
interspersed with observations of the phase calibrator J2357-5311.
This bright quasar was also observed for bandpass and flux
calibration.  The observations utilised 35 antennas with baselines of
$20-650\m$.  The frequency division correlator mode was used for the
spectral line observation with a $1.875\GHz$ bandwidth and frequency
resolution of $7812\kHz$.  The velocity channels were binned to a
resolution of $12\kmps$ for the subsequent analysis.  Based on optical
spectroscopy of the central galaxy, we use a velocity center at a
redshift $z=0.596$, which also corresponds to the velocity center of
the molecular emission peak.  We note that optical IFU observations have
revealed a very dynamic environment in the ionized gas and bulk
motions could produce a systematic offset in the gas velocities with
respect to the gravitational potential of the BCG (\citealt{McDonald14}).


The observations were calibrated in \textsc{casa} version
4.3.1 (\citealt{McMullin07}) using the ALMA pipeline reduction scripts.  Continuum-subtracted data cubes were generated using \textsc{clean} and \textsc{uvcontsub}.
Additional self-calibration did not produce significant improvement in
the image quality.  Different weightings were explored to determine
the optimum for imaging.  Natural weighting detected the extended
filaments at the highest signal-to-noise but no major differences
could be discerned between the various weightings due to the good
\textit{uv} coverage.  The final data cube used natural weighting and
had a synthesized beam of $0.60\asec\times0.56\asec$ with
P.A. $=-37.9\deg$.  The rms noise in the line-free channels was
$0.3\mJypbm$ at CO(3-2) for $12\kmps$ channels.  An image of the
continuum emission with an rms noise of $0.019\mJypbm$ was generated
by combining spectral channels from all four basebands that were free
of line emission.  The continuum image was produced using natural
weighting and the synthesized beam was $0.59\asec\times0.53\asec$ with
P.A. $=-48.7\deg$.

\begin{table*}
\begin{minipage}{\textwidth}
\caption{Fit parameters for the total CO(3-2) spectrum shown in Fig. \ref{fig:totalspec} corrected for primary beam response.}
\begin{center}
\begin{tabular}{l c c c c c c}
\hline
Region & $\chi^2$/dof & Component & Integrated intensity & Peak & FWHM & Velocity shift \\
 & & & (Jy{\thinspace}km/s) & (mJy) & (km/s) & (km/s) \\
\hline
Total & 198/141 & 1 & $5.3\pm1.0$ & $11.0\pm1.0$ & $450\pm80$ & $20\pm50$ \\
 & & 2 & $2.3\pm1.0$  & $8.5\pm3.0$ & $260\pm70$ & $310\pm20$ \\
 & & 3 & $1.3\pm0.4$  & $4.9\pm0.6$ & $250\pm60$ & $630\pm30$ \\
\hline
\end{tabular}
\end{center}
\label{tab:fits}
\end{minipage}
\end{table*}

\section{Results}

\subsection{Molecular gas morphology}
\label{sec:morph}

The CO(3-2) molecular line emission peaks at the galaxy center, offset
by $\sim0.3\asec$ to the W of the radio nucleus
(Fig. \ref{fig:totalspec} left).  The central molecular gas peak
extends along a NE to SW axis across the nucleus.  Two filaments
extend $3-4\asec$ ($20-27\kpc$) to the NW and SE of the central
emission peak.  The emission also extends for several arcsec as a
broader structure to the S of the nucleus.  Fig. \ref{fig:totalspec}
(right) shows the continuum-subtracted total CO(3-2) spectral line
profile extracted from a $6\arcsec\times6\arcsec$ region centred on the
nucleus and covering all extended emission.  The line profile
is very broad, covering $\sim1000\kmps$, and consists of multiple
velocity components.  The total spectrum was fitted with three
Gaussian components using the package \textsc{mpfit}
(\citealt{Markwardt09}).  The brightest velocity component is centred
on the galaxy's systemic velocity and has the largest FWHM of
$450\pm80\kmps$.  At least two additional velocity components are
redshifted to $310\pm20\kmps$ and $620\pm30\kmps$ and have
significantly lower FWHMs of $\sim250\kmps$.  The best fit results,
corrected for primary beam response and instrumental broadening, are
shown in Table \ref{tab:fits}.

The most luminous redshifted component covering the velocities from
$\sim0$ to $\sim480\kmps$ corresponds to the most extended emission
(Fig. \ref{fig:filimg}).  The remaining molecular gas at $>500\km/s$
and $<0\kmps$ lies within $1.5\asec$ of the nucleus.
Fig. \ref{fig:filimg} clearly shows that this velocity component
traces the most extended emission from the NW and SE filaments and the
third filament to the S.  The filament widths are unresolved
and may consist of many individual strands (\citealt{Fabian08}).  Each
filament coincides with regions of bright ionized gas emission, dust
and filamentary star formation previously detected in optical and UV
observations (Fig. \ref{fig:filimg} right;
\citealt{McDonald13,McDonald14}).


The filaments are draped around the rims of two large X-ray cavities
(Fig. \ref{fig:filimg}, left) detected in deep \textit{Chandra} X-ray
observations (\citealt{McDonald15}).  These cavities are each
$9-15\kpc$ across and centered at a radius of $\sim17\kpc$ and were
created as radio bubbles inflated by the AGN displaced the cluster
gas.  The SE and S filaments encase the inner half of the
southern X-ray cavity, which is larger and has the greater cavity
power of the two.  The NW filament lies along the W edge of the
northern X-ray cavity but no significant counterpart is detected along
the E edge.  The filaments may form part of a patchy thin shell
surrounding the inner half of each bubble.  This molecular shell would
be brightest at the largest projected distance around the bubble,
which could explain the remarkable coincidence between the filaments
and the bubble edges and the limited amount of molecular gas projected
across either bubble.



The total CO(3-2) intensity of $8.9\pm1.5\Jykmps$ is almost a factor
of 2 greater than the integrated intensity measured by the SMA of
$5.3\pm1.4\Jykmps$ (\citealt{McDonald14}).  The SMA observation was
taken at very low elevation, is only modestly significant and likely
affected by a substantial continuum subtraction uncertainty, making it
particularly difficult to calibrate.  The measured FWHM of
$\sim400\kmps$ from the SMA observation is also significantly less than
determined the FWHM of $\sim670\pm20\kmps$ for a single component
spectral fit to the ALMA total spectrum.  This discrepancy in the
total CO(3-2) flux could therefore be due to uncertainty in the SMA
continuum subtraction.

\subsection{Velocity structure}
\label{sec:velocity}

\begin{figure*}
\begin{minipage}{\textwidth}
\centering
\includegraphics[width=0.45\columnwidth]{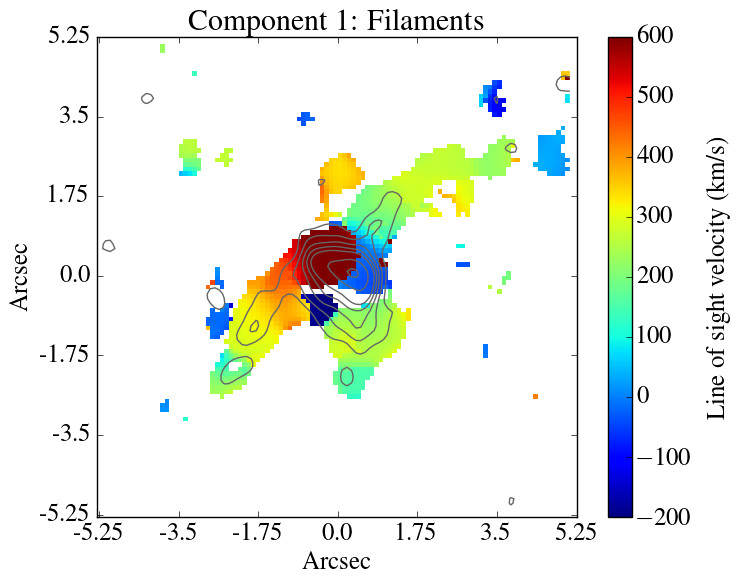}
\includegraphics[width=0.45\columnwidth]{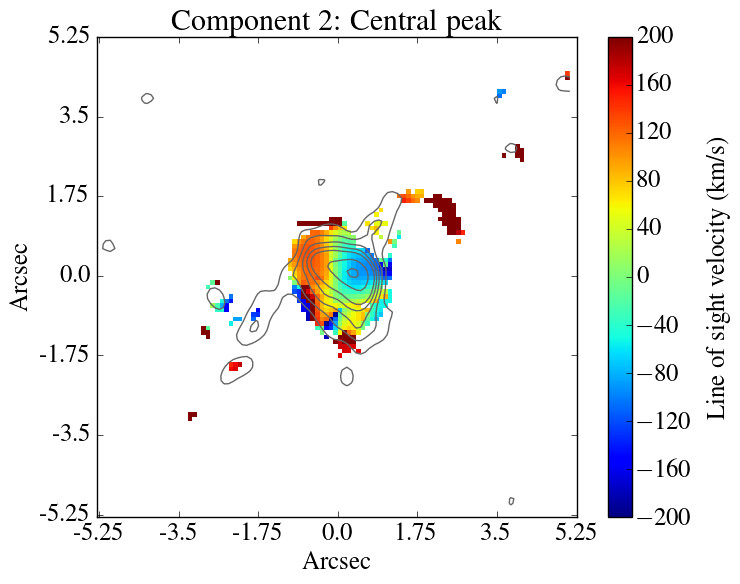}
\caption{Velocity line centre for each component.  Spectral fitting reveals that the molecular gas structure has two distinct velocity components.  The first component (left) traces the filaments and has a narrow FWHM$\sim100-200\kmps$ and smooth velocity gradients along their lengths.  The second component (right) corresponds to the central gas peak and has a much higher FWHM$\sim300-550\kmps$, lower velocities and a gradient E-W across the nucleus.  The contours correspond to Fig. \ref{fig:totalspec} (left)} 
\label{fig:velmaps}
\end{minipage}
\end{figure*}

\begin{figure*}
\begin{minipage}{\textwidth}
\centering
\includegraphics[width=0.45\columnwidth]{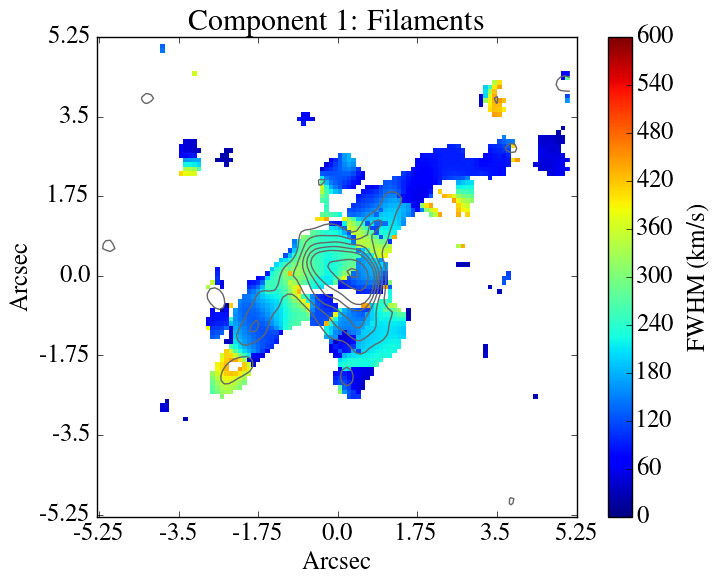}
\includegraphics[width=0.45\columnwidth]{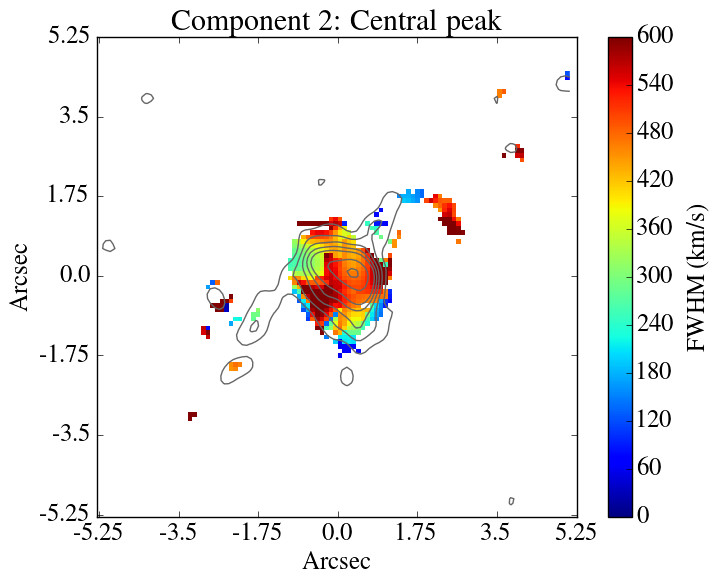}
\caption{Same as Fig. \ref{fig:velmaps} but showing the FWHM of each velocity component.}
\label{fig:fwhmmaps}
\end{minipage}
\end{figure*}


The velocity structure of the molecular gas was mapped by extracting
spectra in synthesized beam-sized regions each centred on each spatial
pixel in the ALMA cube.  The extracted spectra were fitted with
\textsc{mpfit} using one, two or three Gaussian components.  We
required at least $3\sigma$ significance for the detection of an
emission line in each region based on 1000 Monte Carlo simulations of
the spectrum.  Figs. \ref{fig:velmaps} and \ref{fig:fwhmmaps} show the
maps of the line centre and FWHM that were generated for each best-fit
velocity component.  These maps show two distinct components in the
molecular gas: the extended filaments characterized by smooth
velocity gradients and narrow FWHM$=100-200\kmps$ and a compact
central gas peak with much broader FWHM$=300-550\kmps$.



All three extended filaments have ordered velocity gradients along
their lengths and low FWHM$<250\kmps$ decreasing to $<150\kmps$ at the
largest radii in the NW and S filaments.  The smooth velocity
gradients and narrow FWHM over the length of each extended structure
reveal a steady, ordered flow of molecular gas around and beneath the
radio bubbles.  The velocities at the furthest extent of each filament
are similar at $\sim250\kmps$ and increase towards the galaxy nucleus
with the highest velocities at the smallest radii.  In the SE
filament, the velocity increases with decreasing radius to
$\sim600\kmps$ to the E of the nucleus.  The steady velocity gradient
suggests that the SE filament forms a continuous structure to the
galaxy centre.  The velocity gradient along the NW filament, from
$+280\kmps$ at $2.3\asec$ radius ($16\kpc$) to $0\kmps$ at the nucleus,
also indicates a separate, continuous structure.  The S filament is
fainter, shorter and wider and has a shallower velocity gradient.  The
FWHM is below $100\kmps$ at $2.5\asec$ radius ($17\kpc$) but quickly
broadens to $>200\kmps$ with decreasing radius indicating that the S
filament is disrupted towards the galaxy centre.  The SE and NW
filaments meet in projection at the nucleus.  The velocity
differential is large across the nucleus at $+600\kmps$ to $0\kmps$
with no evidence for disruption at this resolution and no significant
increase in FWHM of the velocity components for each filament.
Collisions between the gas clouds in these two ordered filaments may
produce a sharp transition in velocity to the second observed velocity
component.



The central compact molecular gas peak forms a separate velocity
structure from the extended filaments with a much higher
FWHM$=300-550\kmps$ and a E-W velocity gradient from $+120\kmps$ to
$-80\kmps$ across the nucleus (Fig. \ref{fig:velmaps}).  The velocity
gradient across the nucleus appears to lie perpendicular to the
velocity gradients along the NW and SE filaments.  The FWHM peaks
along the projected intersection of the two filaments.  The increase
in FWHM could indicate variation in the orientation of the filaments
at the galaxy centre but, although it could be a contributing effect,
this would require a reversal of the velocity gradient in a similar
length of filament oriented along the line of sight at the galaxy
centre.  It is more likely that the FWHM is intrinsically higher in
the central molecular gas peak.  The E-W velocity gradient centred on
the nucleus is consistent with ordered motion or rotation about the
AGN within a radius of $\sim1\asec$ ($7\kpc$) but higher spatial
resolution observations are required to determine if this corresponds
to a disk.

The strong velocity gradients in the molecular gas are comparable to
those observed for the warm ionized gas (\citealt{McDonald14}).  The
ionized gas, traced by the \textsc{[o~iii]}$\lambda$4959, 5007
doublet, shows a relatively smooth gradient from $+700\kmps$ to the SE
of the nucleus decreasing to $-400\kmps$ to the NW.  The velocity is
$\sim0\kmps$ at the peak of the \textsc{[o~ii]} emission around the
nucleus.  The low velocities of $+200\kmps$ to $-200\kmps$ at the
centre correspond to the bright central peak and are consistent with
the gas velocities around the nucleus in the molecular gas.  The high
velocities in the ionized gas to the SE and NW are comparable to the
bright, innermost regions of the SE and NW filaments.  Although the
ionized gas velocities appear to decrease at larger radii, which
corresponds to the outer regions of the filaments, the emission is
faint and extends beyond the field of view (\citealt{McDonald14}).




The fraction of the total CO(3-2) flux in each filament was estimated
by using the velocity structure to separate the filaments from the
compact central emission peak.  Using a spectrum extracted from a
region covering the central peak, we determined the integrated flux in
each filament based on the best-fit model for their distinct velocity
structures.  This was added to the flux determined from conservative
regions covering only the extended structure of each filament.  The
low velocity structure at large radius in each filament could not be
easily spectrally separated from the low velocities of the gas across
the nucleus.  Therefore it was not possible to use a purely spectral
or purely spatial separation of the extended and compact structures.
From this hybrid technique, we estimate that the SE filament contains
$\sim25\%$ of the total flux and the NW filament contains $\sim15\%$.
The extent of the S filament is particularly uncertain.  Based on the
assumption that all the emission to the S of the nucleus with a FWHM
below $250\kmps$ is associated with the S filament then it contains
$\sim10-15\%$ of the total flux.


In summary, roughly half of the total flux lies in three extended
filaments which have ordered velocity gradients along their lengths
and low FWHM$<250\kmps$.  The velocities at the furthest extent of
each filament are similar at $\sim250\kmps$ and increase towards the
galaxy nucleus with the highest velocities at the smallest radii.  The
central compact emission peak forms a separate velocity structure with
a much higher FWHM$=300-550\kmps$ and a velocity gradient lying
perpendicular to the filaments from $+120\kmps$ to $-80\kmps$ across
the nucleus.  The velocity structure is consistent with ordered motion
around the nucleus but higher spatial resolution observations would be
required to determine if this is a disk.

\subsection{Molecular gas mass}
\label{sec:mass}

By assuming a CO-to-H$_2$ conversion factor $X_{\mathrm{CO}}$ and a
line ratio CO(3-2)/CO(1-0) $\sim0.8$ (\citealt{Edge01,Russell16}), the total
molecular gas mass can be inferred from the integrated CO intensity:

\begin{equation}
M_{\mathrm{mol}}=1.05\times10^4X_{\mathrm{CO}}\left(\frac{1}{1+z}\right)\left(\frac{S_{\mathrm{CO}}\Delta\nu}{\Jykmps}\right)\left(\frac{D_{\mathrm{L}}}{\Mpc}\right)^2\Msun,
\end{equation}

\noindent where $D_{\mathrm{L}}$ is the luminosity distance, $z$ is
the redshift of the BCG and $S_{\mathrm{CO}}\Delta\nu$ is the
integrated CO(1-0) intensity.  However, the molecular gas mass is
particularly sensitive to the CO-to-H$_2$ conversion factor, which is
quite uncertain and not expected or observed to be universal (see
\citealt{Bolatto13} for a review).  Previous studies of BCGs in cool core
clusters (\citealt{Edge01,Salome03,Russell14,McNamara14}) have used the
Galactic value $X_{\mathrm{CO}}=2\times10^{20}\COtoH$
(\citealt{Solomon87,Solomon05}).  However in intense starburst galaxies
and ULIRGs such as the BCG in the Phoenix cluster, the molecular gas
exists at higher densities and temperatures producing an extended warm
gas phase with a much higher column density than a quiescent system.
Under these conditions, the CO emission is more luminous and
$X_{\mathrm{CO}}$ should be lowered (eg. \citealt{Downes93,Iono07,Aravena16}).  The
high star formation density of $5\Msunpyrpsqkpc$ within the central
$10\kpc$ (\citealt{McDonald14}), warm dust temperature of $87\K$ and
large FWHM in the galaxy center suggest that a lower $X_{\mathrm{CO}}$
is appropriate for the central molecular gas structure and potentially
also for the filaments.  We therefore assume
$X_{\mathrm{CO}}=0.4\times10^{20}\COtoH$ (\citealt{Downes98}) but note
that \textit{Chandra} observations measure subsolar metallicity of
$0.5\Zsun$ in the surrounding ICM.  Low metal abundance likely will
boost $X_{\mathrm{CO}}$ over our assumed value, unless the cool gas in
the filaments has an increased metallicity over the ambient medium
(eg. \citealt{Panagoulia13}).

For the integrated CO(3-2) intensity of $8.9\pm1.5\Jykmps$, the total
molecular gas mass is $2.1\pm0.3\times10^{10}\Msun$.  As discussed in
section \ref{sec:morph}, the integrated intensity is almost a factor of two
higher than that found by the earlier SMA observation
(\citealt{McDonald14}), which is likely due to uncertainty in the
continuum subtraction for the SMA result.  Note also that
\citet{McDonald14} assume a line ratio CO(3-2)/CO(1-0) $\sim0.5$,
which produces a similar molecular gas mass despite the difference in
integrated intensity.  The central molecular gas peak in the Phoenix
cluster accounts for $\sim50\%$ of the total CO flux and therefore has
a molecular gas mass of $1.0\pm0.2\times10^{10}\Msun$.  The SE, NW and
S filaments contain $\sim0.5\times10^{10}\Msun$,
$\sim0.3\times10^{10}\Msun$ and $\sim0.3\times10^{10}\Msun$,
respectively.  The uncertainty on the $X_{\mathrm{CO}}$ factor
increases the uncertainty on the molecular gas masses to roughly a
factor of a few.  This estimate of the molecular gas mass also appears
low when compared with correlations with the H$\alpha$ luminosity
(\citealt{Salome03}) and the dust-to-gas ratio of $\sim20$
(\citealt{McDonald12}).  However, our conclusions are not
qualitatively altered by the estimated uncertainty.


\subsection{AGN continuum}

An unresolved central continuum source was detected at RA
23:44:43.902, Dec -42:43:12.53 with a flux of $2.5\pm0.1\mJy$ at
$225.09\GHz$.  The position and flux are
consistent with the SMA continuum detection at $3\mJy$
(\citealt{McDonald14}\footnote{Note that the lower SMA continuum flux
  given in \citealt{McDonald14} is a typographical error.}).  The ALMA
continuum image does not reveal any extended emission due to star
formation.  The continuum source is coincident with radio and hard X-ray point source emission.
Low frequency radio observations from the SUMSS and ATCA
archives suggest a synchrotron continuum slope of
$S\propto\nu^{-1.35}$ and therefore we expect synchrotron emission
from the AGN of $\sim0.04\mJy$ at $220\GHz$ (\citealt{McDonald14}).
The observed point source flux is therefore consistent with a combination of synchrotron
emission and dust emission from the SMBH's immediate environment.


\section{Discussion}

\begin{figure*}
\begin{minipage}{\textwidth}
\centering
\raisebox{0.1cm}{\includegraphics[width=0.57\columnwidth]{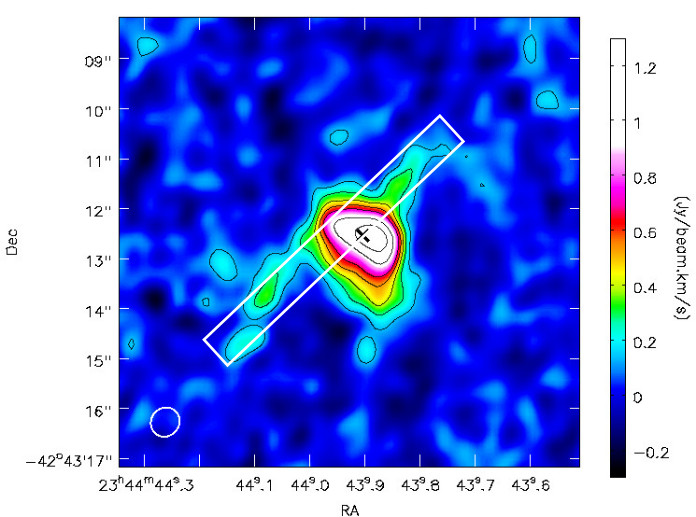}}
\includegraphics[width=0.42\columnwidth]{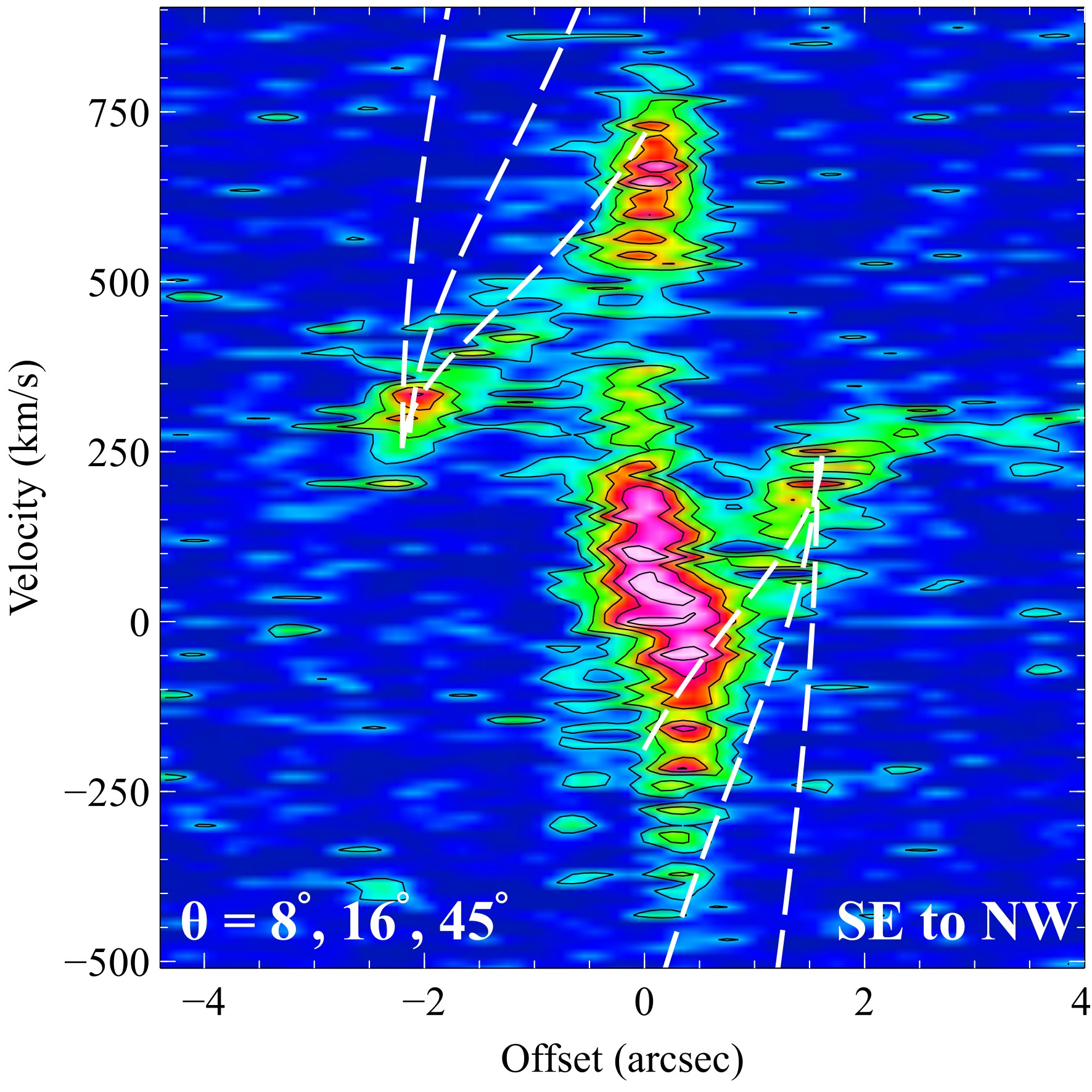}
\caption{Left: Integrated intensity map for velocities $-430$ to $+600\kmps$ covering both the central gas peak and the extended filaments.  Contours are $2\sigma$, $4\sigma$, $6\sigma$, $8\sigma$, $10\sigma$, $15\sigma$, $20\sigma$, $25\sigma$ and $30\sigma$, where $\sigma=0.065\Jypbmkmps$.  The white box shows the extraction region for the position-velocity diagram.  Right: Position-velocity diagram for the SE to NW axis along the two brightest filaments summed over roughly the width of the synthesized beam.  Model velocity profiles are shown by the dashed lines for gravitational free-fall with inclinations 8, 16 and $45^{\circ}$.  For the SE filament, initial radii are 15, 15 and $21\kpc$, respectively.  For the NW filament, initial radii are 11, 11 and $15\kpc$, respectively.  Inclinations $<15^{\circ}$ are required to match the velocity gradient of both filaments.}
\label{fig:filpv}
\end{minipage}
\end{figure*}


In the central galaxy of the Phoenix cluster, massive molecular gas
filaments form dense, cold rims around both of the inner X-ray
cavities, where hot gas has been displaced by radio jet activity.
These observations now clearly demonstrate that the structure of the
largest molecular gas reservoirs located in the most massive galaxies
is shaped by the expansion and trajectory of the radio bubbles.
Previous sub-mm observations of brightest cluster galaxies have
indicated tentative correlations between X-ray cavity axes and the
orientations of molecular gas filaments, including ALMA observations
of Abell 1835 and PKS\,0745-191 (\citealt{McNamara14,Russell16}).
IRAM observations of the nearby Perseus cluster detected molecular gas
coincident with regions of the complex optical emission line nebula,
including several filaments of ionized gas that extend toward radio
bubbles (\citealt{Salome06,Lim08,Salome11}).  These observed direct
interactions between the cold gas, which fuels the starburst and black
hole activity, and the jet-blown bubbles are essential to explain the
observed close regulation of AGN feedback.

The total molecular gas mass of $2.1\pm0.3\times10^{10}\Msun$ (section
\ref{sec:mass}) in the central galaxy of the Phoenix cluster is
substantially higher than that typically found in early-type galaxies
(\citealt{Young11}).  BCGs situated in dense cluster atmospheres with
short radiative cooling times are known to preferentially host cold
molecular gas in excess of several $10^9\Msun$
(\citealt{Edge01,Salome03}).  The molecular gas structures are
observed to be coincident with bright optical emission line nebulae
and the most rapidly cooling X-ray gas
(\citealt{FabianFil03,Salome06}).  In the Phoenix cluster, the
molecular filaments are similarly coincident with the brightest, soft
X-ray emission, ionized gas plumes and dust regions.  The X-ray gas
cooling rate measured with XMM-Newton RGS of
$120^{+340}_{-120}\Msunpyr$ (\citealt{Tozzi15}) is consistent with the
observed mass of molecular gas originating in cooling of the hot
atmosphere over roughly the buoyant rise time of the inner and outer bubbles in the Phoenix cluster
($50-120\Myr$, see also \citealt{Russell14,McNamara14}).  An
alternative merger origin for such a substantial mass of molecular gas
would require multiple gas-rich donor galaxies, which are rare at the
centres of rich clusters (eg. \citealt{Young11}).  It is more likely
that the molecular gas originated in gas cooling from the surrounding
hot atmosphere.


The molecular gas clouds have either been subsequently lifted into
extended filaments by the expanding radio bubbles or formed in filaments
via thermal instabilities induced in uplifted low entropy X-ray gas.
Radio jets have been observed to drive significant outflows of
molecular gas from galaxies
(\citealt{Morganti05,Dasyra12,Morganti15}).  For the Phoenix cluster,
\citet{McDonald15} estimated the mechanical jet energy from the work
done displacing the hot gas against the surrounding pressure.  By measuring
the size of the inner cavities and the local gas pressure, the cavity
enthalpy was estimated as $4.4-6.7\times10^{59}\erg$.  Only a few per
cent of this mechanical energy could supply the observed kinetic
energy of the molecular filaments.  However, the coupling mechanism
between dense, molecular clouds and radio bubbles is unclear and this
mechanism would have to be extremely efficient to lift 50\% of the
cold gas into extended filaments.  Volume-filling X-ray gas would be
much easier to lift.  

Outflows of hot X-ray gas, enriched with metals by stellar activity in
the central galaxy, are observed in galaxy clusters as plumes of high
metallicity gas lifted along the jet axis for tens to hundreds of kpc
(\citealt{Simionescu08,Kirkpatrick09}).  Low entropy X-ray gas should
become thermally unstable when lifted to a radius where its cooling
time approaches the infall time
(\citealt{Nulsen86,PizzolatoSoker05,McNamara14,McNamara16}).
Theoretical models further indicate that lifting low entropy gas in
the updraft of rising radio bubbles stimulates condensation of
molecular clouds (\citealt{Li14,Brighenti15,Voit16}).  Therefore, an
infall time that is significantly longer than the free-fall time, will
enhance thermal instabilities and promote the formation of molecular
gas clouds in the bubbles' wakes (\citealt{McNamara16}).

The inner radio bubbles in the Phoenix cluster displace
$\sim3-5\times10^{10}\Msun$ and therefore, by Archimedes' principle,
could lift the $1.0\times10^{10}\Msun$ of gas required to supply the
molecular gas in the filaments.  The similarity in the molecular gas
velocity at large radius, and in the velocity gradients beneath
bubbles with apparently different dimensions, supports this scenario
where the molecular gas cools and decouples from the hot atmosphere
and falls toward the galaxy center.  Whilst the velocity range covers
$\sim1000\kmps$ at the galaxy center, the molecular gas velocities at
the outer tip of each filament, separated by $\sim30\kpc$, are
consistent with $\sim250\kmps$.  Such similar velocities suggest that
this remote molecular gas could be coupled to the hot atmosphere
(\citealt{Hitomi16}), which is moving relative to the BCG.  Bulk
motion of the cluster gas could also explain the bubble asymmetry.
The smoothly increasing gas velocities with decreasing radius along
the NW and SE filaments indicate massive gas flows underneath the radio bubbles.  Although the velocity structure cannot cleanly
distinguish between inflow and outflow, the remarkable similarity in
the molecular gas velocities at large radii suggests the molecular gas could be  decoupling from the hot atmosphere and the smoothly increasing velocities toward the galaxy centre suggest subsequent infall.


The smooth velocity gradients along the SE and NW filaments are shown
clearly in Fig. \ref{fig:filpv}.  
Following \citet{Lim08}, we assume a Hernquist model for the gravitational
potential of the central galaxy (\citealt{Hernquist90}) constrained by
parameters for the total galaxy mass $M$ and effective radius $R_e$.
The velocity acquired by a gas blob that free-falls in this
gravitational potential is given by

\begin{equation}
v(r)=\sqrt{v(r_0)^2 + 2GM\left(\frac{1}{r+a} - \frac{1}{r_0 + a}\right)} ,
\end{equation}

\noindent where $a$ is the scale radius ($R_e\sim1.8153a$), $v(r_0)$
is the initial velocity of the gas blob and $r_0$ is its initial
radius.  \citet{McDonald12} (see also \citealt{McDonald13}) determine an effective radius of
$\sim17\kpc$ from HST imaging in five photometric bands.  The scale
radius is therefore estimated at $9.4\kpc$.  

The total galaxy mass was estimated from \textit{Chandra} observations
assuming hydrostatic equilibrium for the hot X-ray atmosphere in the
gravitational potential.  We used spectra extracted in concentric
annuli (\citealt{McDonald15}) and the \textsc{nfwmass} model
(\citealt{Nulsen10}) in \textsc{xspec} (version 12.9.0;
\citealt{Arnaud96}) to determine the best-fit NFW profile parameters
assuming a spherical, hydrostatic atmosphere.  The best-fit scale
radius $r_{s}=200^{+40}_{-30}\kpc$ and the normalization constant
$\rho_0=7\pm2\times10^6\Msunpcukpc$.  We therefore estimate a galaxy
total mass of $\sim2\times10^{13}\Msun$ within a radius of
$\sim50\kpc$.  For the best-fit NFW profile, the cluster mass within
$r_{500}\sim1.3\Mpc$ is $M_{500}=7\pm4\times10^{14}\Msun$.  This is
consistent with the mass determined from scaling relations with
$Y_{\mathrm{SZ}}$ (\citealt{Williamson11}) and with $Y_{\mathrm{X}}$
(\citealt{McDonald15}).


The remaining free parameters for the Hernquist model are the initial
radius of the gas blob, the inclination of the trajectory to the line
of sight and the initial velocity, where the clouds are coupled to the hot atmosphere.  As discussed in
section \ref{sec:velocity}, the outermost velocities in the filaments
are similar and therefore all models used $250\kmps$.  The initial
radius was selected to match each model to the outermost region of the
appropriate filament.  Fig. \ref{fig:filpv} shows the PV diagram for
the NW to SE axis along the NW and SE filaments with free-fall models
for several inclination angles.  The observed shallow velocity
gradients can only be matched by the lowest inclination angles
$\theta<15^{\circ}$, for which both filaments are oriented close to
the plane of the sky.  Beyond the initial acceleration, the model
velocity gradient does not depend on the initial radius and velocity.
Although the inclination and the total mass are degenerate, the total
mass would have to be overestimated by at least a factor of a few to
allow inclination angles of $>20^{\circ}$.  This would also likely
require the total stellar mass of $3\times10^{12}\Msun$
(\citealt{McDonald12,McDonald13}) to have been significantly
overestimated.  We therefore suggest that such stringent requirements
for the orientations of all three filaments, and similar results from
ALMA observations of PKS\,0745-191 and Abell 1835
(\citealt{McNamara14,Russell16}), demonstrate that the gas velocities
are more likely intrinsically low.  Rather than require that the
observed filaments are all aligned in the plane of the sky, we suggest
that the filament velocities are inconsistent with gravitational
free-fall.  In addition to the effect of inclination, the infalling
gas blobs are slowed, potentially by magnetic tension
(\citealt{Fabian08,Russell16}) or by cloud-cloud collisions within the
central few kpc (\citealt{PizzolatoSoker05,Gaspari15}).



\section{Conclusions}

Half of the $2.1\pm0.3\times10^{10}\Msun$ molecular gas reservoir at
the center of the Phoenix cluster lies in extended filaments draped
around expanding radio bubbles inflated by relativistic jets and
powered by the SMBH.  The filaments have smooth velocity gradients
along their lengths and narrow line widths consistent with massive,
ordered gas flows around the radio bubbles.  Although the velocity
structure alone does not allow us to distinguish cleanly between
inflow or outflow, the massive molecular gas flow is clearly shaped by
the recent radio-jet activity.  The molecular gas may have been
directly lifted in the bubble wakes or formed in situ at
large radius from uplifted low entropy X-ray gas that became thermally
unstable.  
The gas velocities appear too low for the bulk of the cold
gas to escape the galaxy and the gas will eventually fall back toward
the galaxy center to feed the central gas peak.  The observed close
coupling between the radio bubbles and the cold gas is essential
to explain the self-regulation of feedback and understand the
stability of this mechanism in clusters over at least half the age of
the universe (\citealt{McDonald13survey,Ma13,HlavacekLarrondo15}).

\section*{Acknowledgements}

HRR and ACF acknowledge support from ERC Advanced Grant Feedback
340442.  MM acknowledges support by NASA through contracts
HST-GO-13456 (Hubble) and GO4-15122A (Chandra).  BRM acknowledges
support from the Natural Sciences and Engineering Council of Canada
and the Canadian Space Agency Space Science Enhancement Program.  PEJN
acknowledges support from NASA contract NAS8-03060.  BB is supported
by the Fermi Research Alliance, LLC under Contract
No. De-AC02-07CH11359 with the United States Department of Energy.
ACE acknowledges support from STFC grant ST/L00075X/1.  JHL
acknowledges support from the Natural Sciences and Engineering Council
of Canada, the Canada Research Chairs program and the Fonds de
recherche Nature et technologies. CR acknowledges support from the
Australian Research Council's Discovery Projects funding scheme
(DP150103208).  We thank the reviewer for constructive comments and
HRR thanks Adrian Vantyghem for helpful discussions.  This paper makes
use of the following ALMA data: ADS/JAO.ALMA 2013.1.01302.S. ALMA is a
partnership of ESO (representing its member states), NSF (USA) and
NINS (Japan), together with NRC (Canada), NSC and ASIAA (Taiwan), and
KASI (Republic of Korea), in cooperation with the Republic of
Chile. The Joint ALMA Observatory is operated by ESO, AUI/NRAO and
NAOJ.  The scientific results reported in this article are based on
data obtained from the Chandra Data Archive.


\bibliographystyle{aasjournal}  
\bibliography{refs.bib}

\clearpage

\end{document}